\documentclass[12pt]{article}
\usepackage{amsmath,amssymb,mathrsfs}
\usepackage{bm}
\usepackage{cite}

\newcommand{\lb}{\left}
\newcommand{\rb}{\right}
\renewcommand{\to}{\ensuremath{\rightarrow}}
\newcommand{\To}{\ensuremath{\Rightarrow}}

\newcommand{\abs}[1]{{|#1|}}

\newcommand{\sfrac}[2]{{#1/#2}}
\newcommand{\Cmb}[2]{{{}_{#1}\mathrm{C}_{#2}}}

\newcommand{\R}{\ensuremath{\mathbb{R}}}

\newcommand{\Half}{\ensuremath{\frac{1}{2}}}

\newcommand{\Quarter}{\ensuremath{\frac{1}{4}}}

\newcommand{\AdS}{\ensuremath{{\mathrm{AdS}}}}

\newcommand{\SU}[1]{\ensuremath{{\mathrm{SU}\!\left(#1\right)}}}

\newcommand{\SO}[1]{\ensuremath{{\mathrm{SO}\!\left(#1\right)}}}

\newcommand{\lp}{\ensuremath{l_p}}

\newcommand{\del}{{\partial}}

\begin{document}
\begin{titlepage}

\thispagestyle{empty}
\begin{flushright}
arXiv:0805.1405 [hep-th] \\
YITP-08-37 \\
May, 2008
\end{flushright}

\bigskip

\begin{center}
\noindent{\Large
\textbf{Coarse-graining of bubbling geometries\\ and the fuzzball conjecture}}\\
\vspace{2cm}
\noindent{
Noriaki Ogawa\footnote{E-mail: noriaki(at)yukawa.kyoto-u.ac.jp}
and
Seiji Terashima\footnote{E-mail: terasima(at)yukawa.kyoto-u.ac.jp}
}\\

\vspace{.2cm}

 {\it  Yukawa Institute for Theoretical Physics, Kyoto University\\
Kyoto 606-8502, Japan}
\vskip 2em
\bigskip
\end{center}

\begin{abstract}
In the LLM bubbling geometries,
we compute the entropies of black holes
and estimate their ``horizon'' sizes from the fuzzball conjecture,
based on coarse-graining on the gravity side.
The differences of black hole microstates cannot be seen by classical observations.
Conversely, by counting the possible deformations of the geometry
which are not classically detectable,
we can calculate the entropy.
We carry out this method on the black holes of
the LLM bubbling geometries, such as the superstar,
and obtain the same result as was derived
by coarse-graining directly on the CFT (fermion) side.
Second,
by application of this method,
we can estimate the ``horizon'' sizes  
of those black holes,
based on the fuzzball conjecture.
The Bekenstein-Hawking entropy computed from this ``horizon''
agrees with that microscopic entropy above.
This result supports the fuzzball conjecture.
\end{abstract}
\end{titlepage}

\tableofcontents

\section{Introduction}
In string theory, black holes are very interesting and important objects.
They are macroscopic systems with nonvanishing entropies,
and in fact they have large numbers of quantum microstates
which account for the entropies.
This can be shown on the dual CFT side in many cases,
as was first derived in \cite{Strominger:1996sh}.
Even black holes with classically vanishing horizon areas 
have many corresponding microstates,
such as the D1-D5 system.

However, 
on the {\it gravity side},
one can ask how these microstates are coarse-grained
and give the black hole geometry with the horizon
(or the stretched horizon in small black hole case),
which obeys the (generalized) Bekenstein-Hawking area law
\cite{Bekenstein:1973ur,Bekenstein:1974ax,Hawking:1976de,Wald:1993nt}.
On this problem,
one interesting and plausible proposal is
the fuzzball conjecture
\cite{Lunin:2001jy,Lunin:2002qf,Mathur:2002ie,Lunin:2002bj,%
Lunin:2002iz,Giusto:2004xm,Mathur:2005zp,%
Balasubramanian:2005qu,Alday:2006nd,%
Skenderis:2006ah,Kanitscheider:2006zf,Kanitscheider:2007wq,%
Larjo:2007zu,Skenderis:2008qn}.
This conjecture was originally proposed as a resolution of
the information loss paradox \cite{Hawking:1976ra},
and includes the following statements:
\begin{enumerate}
\item
Each microstate of a black hole is approximated
by a supergravity geometry
by taking some appropriate basis of the Hilbert space
\footnote{
In fact, this point includes some subtle problems.
For more details, see \cite{Skenderis:2008qn}.
}.
These geometries are all smooth,
without singularities or horizons.
\item
These geometries are not distinguishable
for a classical observer at a distant point,
who cannot observe the Planck scale physics.
They are distinguishable from one another
only within the region 
corresponding to the inner of the black hole macroscopically,
and the boundary of the region becomes the ``horizon''.
Out of the ``horizon'',
all the microstates are observed as the black hole geometry.
\end{enumerate}
On the D1-D5 system,
the geometries corresponding to the microstates,
called fuzzball solutions,
were constructed in \cite{Lunin:2001jy}
and shown to be smooth in \cite{Lunin:2002iz}.
In \cite{Lunin:2001jy}
they discussed the coarse-graining of these solutions,
and showed that their ``horizon'' from the fuzzball conjecture
leads to the Bekenstein-Hawking entropy expected
from the microstates counting.
In spite of the fruitful results including this,
the D1-D5 fuzzball solutions have problems.
They are indeed smooth, but we can also construct
fuzzball-like singular solutions.
In fact, smooth fuzzball solutions are generated from these singular solutions
by some kind of smearing process.
This fact makes it difficult to understand
why the smooth solutions are more fundamental
and represent semiclassical pure states.

Thus it is natural to ask if
there are any other systems
suitable for the investigation of the fuzzball conjecture or not.
Fortunately we have candidates --- the bubbling geometries \cite{Lin:2004nb}.
The bubbling geometries are asymptotic AdS solutions of 10d/11d supergravities,
which correspond to some operators or states on the CFT side,
based on the AdS/CFT correspondence \cite{Maldacena:1997re}.

In this paper, we will deal with
the 1/2 BPS bubbling geometries on $\AdS_5\times S^5$ background.
Smooth bubbling geometries are naturally regarded as
fundamental and semiclassical states,
which are the microstates of singular geometries,
i.e., black holes
\cite{Myers:2001aq,%
Buchel:2004mc,Balasubramanian:2005mg,%
Suryanarayana:2004ig,Shepard:2005zc,%
Balasubramanian:2006jt,D'Errico:2007jm,%
Shieh:2007xn,Balasubramanian:2007zt}.
Then they play roles of fuzzball solutions.
We discuss the coarse-graining of these ``fuzzball solutions''
on the gravity side, in terms of classical observations.
While a similar approach was attempted in \cite{Balasubramanian:2006jt},
our method is more faithful to the principle of coarse-graining.
This leads to a formula of the leading term of the entropy of singular geometries,
which is the same as the one given on the dual CFT side directly
\cite{D'Errico:2007jm, Balasubramanian:2007zt}.
Next we consider observations at closer points to black holes,
and determine the order of the ``horizon'' size based on the fuzzball conjecture.
Substituting this into the naive black hole geometry,
we get a Bekenstein-Hawking entropy which coincides with the
microscopic entropy above.
An early work in this direction is found in \cite{Shepard:2005zc},
although it was not successful.

The construction of this paper is following.
In the next section we shortly review
the LLM (Lin-Lunin-Maldacena) bubbling geometries \cite{Lin:2004nb}
and black holes among them, especially the superstar
\cite{Behrndt:1998ns, Cvetic:1999xp}.
In section 3
we discuss the coarse-graining of them, deriving the entropy formula.
In section 4
we compute the ``horizon'' size from the fuzzball conjecture and
compare it with the entropy.
In the last section we give some remarks and discussion.

\section{LLM geometries and superstar: review}
In this section,
we shortly review the LLM bubbling geometries
and black holes among them, especially the superstar.

\subsection{LLM bubbling geometries}
In the AdS/CFT correspondence,
there is one-to-one correspondence between a state (geometry)
on the AdS side
and an operator (or a state) on the CFT side.
For the correspondence between
the type IIB theory on $\AdS_5\times S^5$
and $\mathcal{N}=4$ $\SU{N}$ Yang-Mills theory,
the classical geometries on the AdS side
which are dual to 
some kinds of operators on the CFT side are known manifestly.

The most representative examples are the $1/2$ BPS chiral primary operators.
On the CFT side, 
using the state-operator mapping,
they are rewritten as the states of the system of
nonrelativistic (1+1)-dimensional $N$ free fermions
in a harmonic oscillator potential
\cite{Hashimoto:2000zp, Corley:2001zk, Berenstein:2004kk}.
The LLM bubbling geometries \cite{Lin:2004nb}
are the  classical geometries of type IIB supergravity,
corresponding to these states.
They are stationary geometries with $\SO{4}\times\SO{4}$ symmetries,
and the metric and the Ramond-Ramond $5$-form field strength $F^{(5)}$ 
are given as follows:
\begin{subequations}
\label{eq:BubblingSolution}
\begin{align}
\label{eq:BubblingMetric}
ds^2 &= -h^{-2}(dt + V_i dx^i)^2 + h^2 (dy^2 + dx^i dx^i)
+ ye^G d\Omega_{(3)}^2 + ye^{-G}d\tilde{\Omega}_{(3)}^2,\\
F^{(5)} &= F\wedge d\Omega_{(3)} + \tilde{F}\wedge d\tilde{\Omega}_{(3)}, \\
\label{eq:DefOfh}
h^{-2} &= 2y\cosh G, \\
\label{eq:RelationBetweenzandV}
y\del_yV_i &= -\epsilon_{ij}\del_ju,\quad y(\del_iV_j-\del_jV_i)=-\epsilon_{ij}\del_yu,\\
\label{eq:DefOfG}
u &= \Half(1-\tanh G),\\
F&=dB_t\wedge(dt+V)+B_tdV + d\hat{B}, \notag\\*
\tilde{F}&=d\tilde{B}_t\wedge(dt+V)+\tilde{B}_tdV + d\hat{\tilde{B}},\\
B_t &= -\Quarter y^2e^{2G},\quad \tilde{B}_t = -\Quarter y^2e^{-2G},\\
d\hat{B}&=\Quarter y^3 *_3 d\lb(\frac{u-1}{y^2}\rb),\quad
d\hat{\tilde{B}}=\Quarter y^3 *_3 d\lb(\frac{u}{y^2}\rb),
\end{align}
\end{subequations}
where
\begin{subequations}
\label{eq:formofuv}
\begin{align}
  \label{eq:formofu}
  u(x_1,x_2,y) &= \frac{y^2}{\pi}\int_{\R^2}\!\!dx'_1dx'_2\frac{u_0(x'_1,x'_2)}{(\abs{\bm{x} - \bm{x'}}^2 + y^2)^2},\\
  \label{eq:formofv}
  V_i(x_1,x_2,y) &= -\frac{\epsilon_{ij}}{\pi}\int_{\R^2}\!\!dx'_1dx'_2
\frac{u_0(x'_1,x'_2)\,(x_j - x'_j)}{(|\bm{x} - \bm{x'}|^2 + y^2)^2},
\end{align}
\end{subequations}
and $*_3$ is the Hodge dual in the $(x_1,x_2,y)$ space.
Here $i,j=1,2$ and $y\ge0$.
All other gauge fields are vanishing,
and the dilaton and the axion fields are constant.
Notice that in this coordinate system,
$x_i$ and $y$ have dimensions of (length)$^2$.
Since the dilaton field is constant,
there is no distinction between the Einstein frame and the string frame.
Thus lengths measured by this metric are physical.
These geometries are completely determined by the single function
$u_0(x_1,x_2) = u(x_1,x_2,0)$ through $u$ and $V_i$.
When $u_0$ satisfies the following conditions
\begin{subequations}
\begin{align}
{}^\exists R,\; x_1^2+x_2^2 > R^2 \;&\To\; u_0(x_1,x_2) = 0, \\
\int_{\R^2}\!\!dx_1dx_2\,u_0(x_1,x_2) &= \pi L^4,
\end{align}
\end{subequations}
\eqref{eq:BubblingSolution} is asymptotically 
$\AdS_5\times S^5$ with $R_{\AdS}=R_{S^5}=L$.
In the AdS/CFT correspondence
\begin{align}
R_{\AdS} = (4\pi N)^{\Quarter}\lp,
\end{align}
so
\begin{align}
\label{eq:TotalChargeOnGravitySide}
\int_{\R^2}\!\!dx_1dx_2\,u_0(x_1,x_2) &= 4\pi^2\lp^4\,N.
\end{align}

On the dual fermion side,
the $u_0(x_1,x_2)$  corresponds to
the distribution of the fermions on the phase plane $(q,p)$.
A fermion occupies the phase plane area $2\pi\hbar$,
so
\begin{align}
\label{eq:TotalChargeOnFermionSide}
\int_{\R^2}\!\!dx_1dx_2\,u_0(x_1,x_2) &= 2\pi\hbar\,N.
\end{align}
In this paper when we refer to $\hbar$,
it is always that on the fermion side.
Comparing \eqref{eq:TotalChargeOnGravitySide} and
\eqref{eq:TotalChargeOnFermionSide},
we find
\begin{align}
\hbar = 2\pi\lp^4.
\end{align}

\subsection{Smooth geometries}
The bubbling geometry \eqref{eq:BubblingSolution}
 has a causal structure
without closed timelike curves
if and only if
\begin{align}
0 \le u_0(x_1,x_2) \le 1,
\end{align}
is satisfied \cite{Milanesi:2005tp}.
In particular, when
\begin{align}
u_0(x_1,x_2) \in \{0,1\}\quad\text{for}\;{}^\forall(x_1,x_2),
\end{align}
the geometry is smooth, without singularities or horizons.
Otherwise it has naked singularities,
as we will see in the next subsection.

On the fermion side,
``semiclassical states'' consist of fermions each of which is individually localized
within a area of about $2\pi\hbar$ on the phase plane.
Then, $u_0$ corresponding to such a state is approximated by
$N$ droplets, which have $u_0=1$ inside the droplets 
and $u_0=0$ outside those,
and the area of each droplet is $2\pi\hbar$
\footnote{
We could define $u_0$ as some distribution
like Wigner or Husimi distribution functions
\cite{Wigner:1932eb,Husimi:1940}
for a superposition of (largely) different semiclassical states.
However, such a distribution function has no classical meaning and is useless,
because a classical observation on it destroys the superposition.
Thus on the gravity side,
we expect that the bubbling geometries corresponding to such $u_0$
should not be regarded as microstates.
For details, see \cite{Balasubramanian:2007zt}.
}.
This represents a smooth geometry on the gravity side.
By contrast, 
$u_0$ with halfway value between $0$ and $1$ corresponds to
some superposition of semiclassical states on the fermion side,
so the state with such $u_0$ cannot be regarded as a semiclassical state.
From the facts above,
it is very natural to assume that
only the smooth bubbling geometries are semiclassical,
and singular ones are not.

Therefore it is convenient to take a basis of the Hilbert space
which consists of smooth geometries on the gravity side.
We will make use of them as the ``fuzzball solutions'' later.
Take the one particle Hilbert space $\mathcal{H}_1$ on the fermion side.
As is well known, the set of one particle coherent states
localized around a phase lattice point
$(\sqrt{2\pi\hbar}\,m,\sqrt{2\pi\hbar}\,n)$
spans a basis of $\mathcal{H}_1$ \cite{vonNeumann:1955,Bargmann:1971ay}.
Then for the Hilbert space of the $N$ fermions system,
the set of the states with $N$ fermions localized
at different phase lattice points individually,
spans a basis, at least approximately.
This is also an approximately orthogonal basis.

\subsection{Singular geometries}
\label{sec:superstar}
A singular bubbling geometry cannot be interpreted
as a semiclassical pure state,
as was discussed in the previous subsection.
Rather, it should be regarded as a coarse-grained state,
which gives an average description of  many semiclassical microstates.
This is very well coincident with the fact that black holes have
nonvanishing entropies generally.

\subsubsection{Superstar}
The simplest and representative example of singular geometry
is following.
Take the $u_0$ configuration as
\begin{align}
\label{eq:SuperstarU0}
u_0
&=
\begin{cases}
\beta &\pod{r < r_0}\\
0 &\pod{r > r_0},
\end{cases}
\end{align}
where $0<\beta<1$ and
we took the polar coordinates $(r,\phi)$ on the $(x_1,x_2)$ plane.
From \eqref{eq:TotalChargeOnFermionSide},
we see that
$r_0$ and $\beta$ satisfies the following relation
\begin{align}
\pi r_0^2\beta &= 2\pi\hbar N \;\lb(=\pi R_{\AdS}^4\rb).
\end{align}
For this configuration , \eqref{eq:formofuv} leads to
\begin{subequations}
\label{eq:zVofSuperstar}
\begin{align}
u
&= \frac{\beta}{2}\lb(1 - \frac{r^2 - r_0^2 + y^2}{\sqrt{(r^2+r_0^2+y^2)^2-4r^2r_0^2}}\rb), \\
V_\phi
&= -\frac{\beta}{2}\lb(\frac{r^2+y^2+r_0^2}{\sqrt{(r^2+r_0^2+y^2)^2-4r^2r_0^2}}-1\rb),\\
V_r &= 0.
\end{align}
\end{subequations}
Here we perform the
coordinate transformation $(t,y,r) $ to $(\tilde{t},\zeta,\theta)$
\begin{subequations}
\label{eq:CoordinatesOfSuperstar}
\begin{align}
y &= R_{\AdS}\,\zeta\cos\theta,\\
r &= R_{\AdS}^2\sqrt{f(\zeta)}\sin\theta,
\quad
f(\zeta)=\frac{1}{\beta}+\frac{\zeta^2}{R_{\AdS}^2},\\
\tilde{t} &= R_{\AdS}\,t,
\end{align}
\end{subequations}
($0\le\theta\le\frac{\pi}{2}$, $\zeta\ge0$).
The metric in this coordinate system is
\begin{subequations}
\begin{align}
ds^2 &=
-\frac{1}{\sqrt{D}}\lb(\cos^2\theta+D\frac{\zeta^2}{R_{\AdS}^2}\rb)d\tilde{t}^2
 +\frac{2R_{\AdS}}{\sqrt{D}}\sin^2\theta\,dtd\phi
 +\frac{R_{\AdS}^2H}{\sqrt{D}}\sin^2\theta\,d\phi^2
\notag\\&\quad
 +\sqrt{D}\lb( f^{-1}d\zeta^2 + \zeta^2\,d\Omega_{(3)}^2 \rb)
 +R_{\AdS}^2\sqrt{D}\,d\theta^2
 +\frac{R_{\AdS}^2}{\sqrt{D}}\cos^2\theta\,d\tilde{\Omega}_{(3)}^2,\\
D &= \sin^2\theta + H\cos^2\theta,
\quad
H = 1+\lb(\frac{1}{\beta}-1\rb)\frac{R_{\AdS}^2}{\zeta^2}.
\end{align}
\end{subequations}
Furthermore, by the dimensinal reduction of the $S^5$ part 
to go to the 5-dimensional $\mathcal{N}=2$ gauged supergravity,
the metric is described as \cite{Cvetic:1999xp}
\begin{align}
\label{eq:SuperStar}
ds_{(5)}^{2}&=
-H^{-\frac{2}{3}}f\,d\tilde{t}^2 + H^{\frac{1}{3}}\lb(f^{-1}d\zeta^2+\zeta^2d\Omega_{(3)}^2\rb).
\end{align}
This is the AdS-background black hole solution known as the superstar
\cite{Behrndt:1998ns}.

In this form of the superstar geometry,
$\zeta = 0$ is a curvature singularity,
and furthermore, it is a naked singularity without horizon.
However, it is believed that, by the effect of higher derivative terms,
this kind of naked singularity develops a stretched horizon
and hides itself behind \cite{Gubser:2000nd}.
These properties of the superstar
are very general for the singular geometries (i.e., black holes)
in this sector.

\section{Observing the LLM geometries}
In the case of ordinary matter,
e.g., gas in a box,
it has a nonvanishing entropy because we cannot distinguish the
microstates.
In other words,
the entropy corresponds to the number of the states which are not
distinguishable from one another by macroscopic observations.

This principle can also be straightforwardly applied to black holes.
In the case of the LLM bubbling geometries,
we know the complete set of the semiclassical microstates in this whole sector.
Thus we can carry out this fundamental method in practice,
to determine the set of the microstates of a black hole and
calculate the entropy.

A similar approach was attempted in \cite{Balasubramanian:2006jt}.
However the basis used in it was the set of the Fock states on the fermion side,
as well as they assumed a certain ensemble by hand.
By the nature of this method, we need not, and should not,
assume any certain ensemble.
It is, on the contrary, automatically determined by the observation.
Furhtermore, 
in order to discuss in terms of macroscopic or classical observation,
we have to adopt a set of semiclassical states as the basis.
It corresponds to the coherent basis on the fermion side.

\subsection{Small differences of geometries}
We take two similar configurations $u_0(x_1,x_2)$ and $u_0'(x_1,x_2)$,
where
\begin{align}i
u'_0 = u_0 + \delta u_0.
\end{align}
In order that 
the corresponding geometries
have same asymptotic AdS radii,
the numbers of fermions, $N$, should be same, thus we require
\begin{align}
\int_{\R^2}\!\!dx_1dx_2\,\delta u_0 = 0.
\end{align}
Under this condition,
the leading source of the differences $\delta u$ and $\delta V_i$ is
the dipole moment of $\delta u_0$.
Since the contributions of the higher multipole moments decrease more
rapidly 
for long distance,
they are expected to be negligible for typical configurations.
So we approximate the $\delta u_0$ as
\begin{subequations}
\begin{align}
\delta u_0(x_1,x_2)
&= 2\pi\hbar n\lb\{\delta^2(\bm{x}-\bm{\xi}) - \delta^2(\bm{x}-\bm{\eta})\rb\},\\
l &= \abs{\bm{\xi}-\bm{\eta}},
\end{align}
\end{subequations}
and define the dipole moment $Q$ of $\delta u_0$ as
\begin{align}
Q &= n\cdot\frac{l}{\sqrt{2\pi\hbar}}.
\end{align}
Later we will estimate the dipole moment Q for the differences
between typical geometries.

We observe these geometries at $(x_1,x_2,y)$ with
\begin{align}
\rho \sim \alpha^2 R_{\AdS}^2,
\end{align}
where
$\rho^2 = x_1^2 + x_2^2 + y^2$,
and
$\alpha$ is some dimensionless constant.
First, we assume that $\rho$ is of macroscopic size,
i.e.,
\begin{align}
\alpha \gtrsim 1.
\end{align}
In this case, generically
\begin{align}
x_1\sim x_2\sim y \sim \rho,
\end{align}
and we will assume it below.

From \eqref{eq:formofu},
the difference $\delta u(x_1,x_2,y)$ of $u(x_1,x_2,y)$ is
\begin{align}
\label{eq:DeltaU}
\delta u
&=  \frac{y^2}{\pi}\int_{\R^2}\!\!dx'_1dx'_2\frac{\delta u_0(x'_1,x'_2)}{(\abs{\bm{x} - \bm{x'}}^2 + y^2)^2} \notag\\
&= 2\hbar ny^2
   \lb\{\frac{1}{(\abs{\bm{x} - \bm{\xi}}^2 + y^2)^2}
        - \frac{1}{(|\bm{x} - \bm{\eta}|^2 + y^2)^2}\rb\} \notag\\
&\approx
  8\hbar ny^2
  \frac{(\bm{\xi}-\bm{\eta})\cdot\bm{x}}{\rho^6}\notag\\
&\sim
  Q\,\lb(\frac{\lp^2}{\rho}\rb)^3 \notag\\
&\sim
  Q\,\alpha^{-6}N^{-\frac{3}{2}}.
\end{align}
Similarly, we see that
\begin{align}
\label{eq:DeltaV}
\delta V_i
&\sim \frac{Q}{\rho}\lb(\frac{\lp^2}{\rho}\rb)^3 \notag\\
&\sim \frac{Q}{{R_{\AdS}^2}}\,\alpha^{-8}N^{-\frac{3}{2}}.
\end{align}

\subsection{Classical observables}
Along our strategy,
we have to estimate
how large differences of $u$ and $V_i$ are detectable for a classical observer.
We assume that one can only measure physical quantities
up to the UV scale $\lambda$ and the IR scale $\Lambda$,
where
\footnote{
We will use only the ratio $\sfrac{\lambda}{\Lambda}$ below.
}
\begin{subequations}
\begin{align}
\lambda &\sim \lp,\\
\Lambda &\sim \sqrt{\rho} \sim \alpha N^{\Quarter}\lp.
\end{align}
\end{subequations}
This is a similar assumption as was used in \cite{Balasubramanian:2006jt}.

The values of
$u$ and $V_i$ themselves are not observable quantities.
We have to measure the geometry, and calculate $u$ and $V_i$ from it.
We can determine the elements of metric by measuring distances.
Due to the limitation of classical observations noted above,
the measured distance is shorter than $\Lambda$,
and includes an error comparable to $\lambda$.
So, perturbations smaller than the original value times
$\frac{\lambda}{\Lambda}\sim \alpha^{-1} N^{-\Quarter}$,
are not detectable.

The magnitudes of $u$ and $V_i$ are
\begin{subequations}
\label{eq:MagnitudeOfUV}
\begin{align}
u
&\sim y^2 \frac{N\hbar}{\rho^4} 
\sim \alpha^{-4},\\
V_i
&\sim \frac{N\hbar\cdot x_j}{\rho^4} 
\sim \frac{1}{R_{\AdS}^2}\alpha^{-6}.
\end{align}
\end{subequations}
From this, one can easily see that
the variation $\delta g$ of any nonzero element $g$ of the metric
\eqref{eq:BubblingSolution} satisfies
\begin{align}
\frac{\delta g}{g}
&\sim \frac{\delta u}{u} + \frac{\delta V_1}{V_1} + \frac{\delta V_2}{V_2} \notag\\
&\sim Q\,\alpha^{-2}N^{-\frac{3}{2}}.
\end{align}
So, the variation of the geometry due to $\delta u_0$ is
detectable for a classical observer,
when $Q$ satisfies
\begin{align}
\label{eq:DetectableCondition}
Q\,\alpha^{-2}N^{-\frac{3}{2}} &\gtrsim \frac{\lambda}{\Lambda}, \notag\\
\therefore\quad
Q &\gtrsim \alpha N^{\frac{5}{4}}.
\end{align}
This result means that
we can make more precise observations
when we are near to the origin.
So we assume that $\alpha \sim 1$ in the next subsection.

\subsection{Entropy of geometries}
In \cite{D'Errico:2007jm,Balasubramanian:2007zt},
a way of coarse-graining the LLM geometries and
calculating the entropy
was proposed,
by directly dealing with the $u_0$ configuration on the fermion side
(while similar analysises were made in \cite{Shieh:2007xn}
for the Lin-Maldacena bubbling geometries and
for the D1-D5 fuzzball geometries in \cite{Larjo:2007zu} ).
Now using the result \eqref{eq:DetectableCondition},
we can rederive their entropy formula,
in terms of classical observation on the gravity side.

For a small but macroscopic (finite at $N\to\infty$) difference $\delta u_0$,
the corresponding $n$ and $l$ are
\begin{subequations}
\begin{align}
n &\sim N,\\
l &\sim N^{\Half}\sqrt{2\pi\hbar},
\end{align}
\end{subequations}
so
\begin{align}
Q\sim N^{\frac{3}{2}} \gg N^{\frac{5}{4}}.
\end{align}
Then such a difference is indeed classically detectable.
While this was assumed in the fermion side approaches above,
we have now successfully derived it on the gravity side.

The next question is the detection bound of smaller differences.
We can detect macroscopic differences of $u_0$ as above,
however we cannot detect sufficiently small ones
--- where is the threshold\,?

As was done in \cite{Balasubramanian:2007zt},
we divide the $(x_1,x_2)$ plane into small regions
with area of order $2\pi\hbar\,M$,
and deform $u_0$ so that the mean value of $u_0$ in each small region
is invariant.
Then we write $\delta u_0 = \sum_k \delta u_0^k$,
in which $\delta u_0^k$ has a support in the region $k$.
Such $\delta u_0^k$ can be approximated by some dipole moment with
\begin{subequations}
\begin{align}
n_k &\sim M,\\
l_k &\sim M^{\Half}\sqrt{2\pi\hbar},
\end{align}
so
\begin{align}
Q_k \sim M^{\frac{3}{2}}.
\end{align}
\end{subequations}
The number of the regions is $\sim \sfrac{N}{M}$,
and typically, each $Q_k$ has a random direction, different from one another.
So the magnitude of the total dipole moment $Q$ scales
as the square root of the number of the regions, $\sim \sqrt{\sfrac{N}{M}}$,
similarly as the traveling distance of a random walk.
Then we see that
\begin{align}
\label{eq:QofSmallDeform}
Q
&\sim M^{\frac{3}{2}} \cdot \sqrt{\frac{N}{M}} \notag\\
&\sim M\sqrt{N}.
\end{align}
Comparing \eqref{eq:DetectableCondition} and \eqref{eq:QofSmallDeform},
\begin{align}
M \gtrsim N^{\frac{3}{4}}
\end{align}
is necessary for classical detection.
In \cite{Balasubramanian:2007zt},
$M$ was assumed as a large number, but not specified.
Now we find that the suitable magnitude of $M$ is
\begin{align}
\label{eq:MagnitudeOfM}
M \sim N^{\frac{3}{4}},
\end{align}
in our setting.

Let us compute the entropy,
assuming that we cannot detect deformations within the regions of
some size $M'$.
In each region $k$, the number of the configurations of $u_0$
with the mean value $u_0^k$ fixed is
\begin{align}
\Cmb{M'}{M'u^k} \sim \exp\lb[ -M'\lb\{ u_0^k\log u_0^k + (1-u_0^k)\log(1-u_0^k) \rb\}\rb],
\end{align}
therefore the total entropy is
\begin{align}
\label{eq:EntropyFormula}
S
&\approx -M' \sum_k \lb\{ u_0^k\log u_0^k + (1-u_0^k)\log(1-u_0^k) \rb\} \notag\\
&\approx -\int\frac{dx_1dx_2}{2\pi\hbar}\lb\{\bar{u}_0\log\bar{u}_0 + (1-\bar{u}_0)\log(1-\bar{u}_0)\rb\},
\end{align}
where $\bar{u}_0(x_1,x_2)$ is the mean value of $u_0$ around $(x_1,x_2)$.
This formula of the leading term of the entorpy is valid
as long as $1 \ll M' \ll N$.
Since \eqref{eq:MagnitudeOfM} satisfies this condition,
the entropy of the geometry is given by \eqref{eq:EntropyFormula}.

\section{Horizon size from fuzzball conjecture}
\label{sec:HorizonSize}
Now we consider the case of smaller $\alpha$.
According to the fuzzball conjecture,
the horizon of a black hole
is the surface of the region
in which the typical microstates are different from each other.
A similar attempt is found in \cite{Shepard:2005zc}.

In the current case,
it is the region in which the entropy formula \eqref{eq:EntropyFormula}
breakes down.
This means $M'\sim 1$.
The corresponding $\alpha$ is very small,
so the approximations in 
\eqref{eq:DeltaU}, \eqref{eq:DeltaV} and \eqref{eq:MagnitudeOfUV}
are not applicable.

Take $\alpha$ very small.
Note that, in this region, what is important is the magnitude of $y$,
rather than $\rho$.
Here we take 
\begin{align}
y\sim \alpha^2 R_{\AdS}^2.
\end{align}
In this case, the effect on $u(x_1,x_2,y)$ from the dipole moment $Q^{\bm{X}}\sim 1$
(which means $n^{\bm{X}}\,l^{\bm{X}}\sim \lp^2$)
at the point $(X_1, X_2)$ on $y=0$ plane can be computed as
\begin{align}
\delta u^{\bm{X}}
\sim y^2 \frac{\lp^6\,\sqrt{(x_1-X_1)^2+(x_2-X_2)^2}}
{\{ (x_1-X_1)^2+(x_2-X_2)^2+y^2 \}^{3}},
\end{align}
similarly as \eqref{eq:DeltaU}.
Then the total contribution is, similarly as \eqref{eq:QofSmallDeform},
\begin{align}
\delta u
&\sim \lb\{ \int\!\frac{dX_1dX_2}{2\pi\hbar} \lb(\delta u^{\bm{X}}\rb)^2 \rb\}^\Half \notag\\
&\sim \lp^4\,y^2 \lb\{ \int_0^{R_{\AdS}^2}\!rdr\, \frac{r^2}{(r^2 + y^2)^6} \rb\}^\Half \notag\\
&\approx \frac{\lp^4}{y^2} \notag\\
&\sim \alpha^{-4}N^{-1}.
\end{align}
Here $u\sim 1$ because the observer is near the boundary $y=0$,
so the observable condition of this $\delta u$ is
\begin{align}
\alpha^{-4}N^{-1} &\gtrsim \alpha^{-1}N^{-\Quarter}, \notag\\
\therefore\quad
\alpha &\lesssim N^{-\Quarter}.
\end{align}
We can easily get the same condition for $\delta V_i$ similarly.

Thus we find that the position of the ``horizon'' is
\begin{align}
\label{eq:YatHorizon}
y \sim N^{-\Half}R_{\AdS}^2.
\end{align}
This means $y\sim \lp^2$, but this coordinate scale is not physical.
In the case of the superstar,
under the coordinate transformation \eqref{eq:CoordinatesOfSuperstar},
this leads to
\begin{align}
\label{eq:ZetaatHorizon}
\zeta \sim N^{-\Half}R_{\AdS},
\end{align}
which precisely agrees with the result in \cite{D'Errico:2007jm},
estimated by the Bekenstein-Hawking entropy based on the
naive metric of the superstar.
Indeed, the ``Bekenstein-Hawking entropy'' computed as the area of
the surface with \eqref{eq:ZetaatHorizon} of the geometry \eqref{eq:SuperStar},
satisfies $S\sim N$, and
the entropy formula \eqref{eq:EntropyFormula} also means $S\sim N$.
Therefore on the superstar and similar black holes,
we have found that the fuzzball conjecture reproduces
the Bekenstein-Hawking law for the $N$ dependence.
This can be regarded as an evidence for the fuzzball conjecture.

\section{Discussion}
In this paper,
we discussed the coarse-graining of the LLM bubbling geometries
and the ``horizons'' from the fuzzball conjecture.
We have successfully derived the entropy formula of the black holes
in terms of the classical observations,
and determined the size of the stretched horizon.
These two quantities exhibited a good agreement
with the Bekenstein-Hawking law.

Unfortunately, however, here is one difficult problem.
In the region around this stretched horizon,
the curvature is very large,
and so the ``fuzzball geometries''
as well as the singular geometry are not reliable in truth.
This is already pointed out, for example in \cite{Skenderis:2008qn}.
But in spite of the large curvature,
the horizon area is rather large, proportionally to $N$.
Therefore it is presumable that,
in the presence of correction terms,
the horizon size would be altered but invariant in the order.

Resolutions of this difficulty is left as a future problem.
One possible way is to deal with black holes with macroscopic horizons,
in which the higher derivative terms will be negligible.
Applications of our method to bubbling geometries with less
supersymmetries (such as the 1/16 BPS sector) would be interesting,
while there might be some technical difficulties.

\section*{Acknowledgements}
We would like to thank Ryo Suzuki, Yuji Tachikawa and Satoshi Yamaguchi for useful discussions.
S.~T.~is partly supported by
the Japan Ministry of Education, Culture, Sports, Science and Technology. 


\providecommand{\href}[2]{#2}\begingroup\raggedright\endgroup

\end{document}